\documentclass{ws-ijmpd}
\usepackage[super,compress]{cite}
\usepackage[breaklinks]{hyperref}
\hypersetup{colorlinks,urlcolor=black,citecolor=black,linkcolor=black,filecolor=black}
\usepackage{feynmf}

\begin{document}

\markboth{Valery V. Nikulin, Sergey G. Rubin}
{Cosmological baryon/lepton asymmetry in terms of Kaluza-Klein extra dimensions}

%
\catchline{}{}{}{}{}
%

\title{\uppercase{Cosmological baryon/lepton asymmetry\\in terms of Kaluza-Klein extra dimensions}}

\author{Valery V. Nikulin}

\address{Department of Particle Physics, National Research Nuclear University MEPhI\\
Kashirskoe shosse, 31,
Moscow, 115409,
Russia\\
n-valer@yandex.ru}

\author{Sergey G. Rubin}

\address{Department of Particle Physics, National Research Nuclear University MEPhI\\
Kashirskoe shosse, 31,
Moscow, 115409,
Russia\\
sergeirubin@list.ru}

\maketitle

\begin{abstract}
In this article, we discuss the mechanism for generating the lepton/baryon number and its subsequent conservation based on extradimensional evolution. The internal angular momentum of extra space is related to the lepton/baryon charge.
Extra space metric are assumed to be initially asymmetric. Symmetries appear during inflation due to an increase in entropy.
Using the multidimensional field toy model, we show that internal angular momentum can accumulate in fermions, which could explain the lepton/baryon asymmetry of the Universe.
\end{abstract}


\ccode{PACS numbers: 04.50.Cd, 04.50.-h, 11.10.Kk, 11.30.-j, 11.30.Fs, 98.80.Cq}


\section{Introduction}

The observed baryon/lepton asymmetry is one of the most challenging problems nowadays. The same existence of the charged Universe means that the charge was not conserved during some period of evolution \cite{1988ZhETF..94....1D}. It seems probable that the inflation is responsible for the baryon large scale distribution and the baryon asymmetry of the Universe \cite{1991ApJ...379..427Y,KhlopovRubin}. Nevertheless, it is conserved with great accuracy at the present time. Hence, an appropriate symmetry responsible for the charge conservation should appear at a post-inflationary stage.
	
At the same time, the idea of extra space almost inevitably leads to charge non conservation \cite{Rubin:2018ybs, Shaposhnikov2001, Dvali2002}. Indeed, in the framework of multidimensional gravity, the observed low energy symmetries are the consequences of the extra space isometries \cite{Blagojevic}. As was discussed in Ref.~\refcite{Kirillov:2012gy} the measure of symmetric manifolds which are nucleated at the sub-planckian energies is zero. There are no any Killing vectors of nucleated manifolds in the very beginning. Hence, there are no conserved charges at the initial stages.  Charge conservation appears much later, when the extra space geometry stabilizes and acquires appropriate Killing vectors.
Therefore, it should be a period of extra space symmetrization. It is exactly what we need in our attempt to solve the  problem of the lepton/baryon asymmetry.
Modern models of extra dimensions, in particular based on the Kaluza-Klein extra dimensions, are quite diverse \cite{overduin1997kaluza, krasnov2018gravity}. Here, we assume that the baryon/lepton symmetry (asymmetry) is the consequence of the compact extra space symmetry (asymmetry).

As shown earlier in Ref.~\refcite{Rubin:2018ybs} the high energy dynamics of the extra space metric leads to a charge accumulation that is conserved at low energies. Scalar fields can accumulate a substantial amount of such charge which could be a reason for the observable lepton or baryon asymmetry of the Universe. In reality, this asymmetry is maintained by fermions rather than the scalar field. 

In this paper, we elaborate the mechanism of the charge transition from the scalar field to the fermions. In our approach, the charge relates to the internal momentum that concerns both scalar and spinor fields. The particles of the scalar field decay into fermions, and the internal momentum transits from the scalar field to fermions.
Notice that the mechanism proposed in our work differs significantly from the Affleck-Dine mechanism \cite{affleck1985new}, in which a scalar field carrying the baryon number is inserted manually.

It was shown in the Ref.~\refcite{Gogberashvili_2007} that a compact apple-shaped extra space with an angle excess gives rise a fermion triplet in the massless KK-mode. In our work, we base on the fact that the fermion triplet arising in such a geometry has a conserved number --- the angular momentum associated with the rotational symmetry of the extra space. We interpret it as a baryon/lepton number. Here we study and clarify the subject.

\section{Model setup}
Our model is based on the 6-dimensional space $M$, which is the direct product of 4-dim space $M_4$ and 2-dimensional compact extra space $K$. We use the following notation: $X_A$ --- 6-dim coordinates, $x_\alpha$ --- 4-dim coordinates, $y_a$ --- 2-dim extra coordinates, $G_{MN}$ --- 6-dim metric, $g_{\mu\nu}$ --- 4-dim metric and $k_{mn}$ --- 2-dim extra metric. From $M=M_4\times K$ follows that determinant $\sqrt{|G(X)|}=\sqrt{|g(x)|}\sqrt{|k(y)|}$. The structure of compact space is discussed below.

The multidimensional fermion field is other necessary component. As we show below, these fermions could carry the lepton/baryon number. 
Also, we have to introduce the scalar field which is responsible for the accumulation of this number and its interaction with the fermions to transfer the number. Thus, the action consists of the following terms:

\begin{align}\label{Aaction}
&S=S_{\text{grav}}+S_{\text{matter}}\,,\nonumber\\
&S_{\text{matter}}=S_{\Phi}+S_{\Psi}+S_{\text{int}}\,.
\end{align}

The term $S_{\Phi}$ represents the action for the 6-dim scalar field. Its role is the number (the lepton/baryon number) storage at the early period of the Universe evolution.
The term $S_{\Psi}$ describes the fermions acting in the 6-dim space. The term $S_{\text{int}}$ contains scalar-fermion interaction which is responsible for the number transfer to the fermions. 

The lepton/baryon number is accumulated in the classical part of the scalar field during the inflation \cite{nikulin2019inflationary}. After the inflation is finished, this number is transferred into the fermions. The aim of this research is to study this transition in detail. We will only be interested in the material part of the action. The issues of gravitational stability and evolution of the extra space are discussed in various papers \cite{nikulin2020formation,Bronnikov:2020tdo}.

\subsection{6-dimensional fields}

The Lagrangian of the scalar field and its interaction with the fermion field is chosen as simple as possible
\begin{equation}\label{scaction}
S_{\Phi}+S_{\text{int}}=\int d^{6} X \sqrt{|G|}\left[\frac{1}{2}\partial_M\Phi \partial^M\Phi-V(\Phi)+f\Phi\Bar{\Psi}\Psi\right]\,.
\end{equation}
In this work, we do not specify the form of the potential $V(\Phi)$ --- it can be ignored and it is used only in subsection \ref{Higgs}.

Fermions acting in the curved 6-dim space-time are described by action \cite{Blagojevic}
\begin{equation}\label{action}
S_{\Psi}=\int\sqrt{|G|}\,d^{6}X\,\,i \overline{\Psi} h_{\tilde{A}}^{B} \hat\Gamma^{\tilde{A}} \nabla_{B} \Psi\,,
\end{equation}
where $\Gamma^{\tilde{A}}$ is the flat gamma matrices, $G_{AB}$ is the metric and $h_{\tilde{A}}^{B}$ is the frame field. The latter relates to the metric as follows:
\begin{equation}\label{mettet}
G_{A B}=h_{A}^{\widetilde{A}} h_{B}^{\widetilde{B}} \eta_{\widetilde{A} \widetilde{B}}\,.
\end{equation}
We choose the gamma matrices in 6 dimensions (two extra coordinates $y_a$ are denoted as $\theta,\varphi$) as follows
\begin{equation}\label{gamma}
\hat\Gamma_{\nu}=\left(\begin{array}{cc}{\gamma_{\nu}} & {0} \\ {0} & {-\gamma_{\nu}}\end{array}\right), \quad \hat\Gamma_{\theta}=\left(\begin{array}{cc}{0} & {-1} \\ {1} & {0}\end{array}\right), \quad \hat\Gamma_{\varphi}=\left(\begin{array}{cc}{0} & {i} \\ {i} & {0}\end{array}\right)\,.
\end{equation}

The covariant derivative of the fermion field is expressed in terms of the spin connection:
\begin{eqnarray}\label{covder}
\nabla_A\Psi&=&\left(\partial_A+\frac{1}{4}\omega_{A\,\tilde{M}\tilde{N}} \left[\hat\Gamma^{\tilde{M}}\hat\Gamma^{\tilde{N}}\right] \right)\Psi\,,
\end{eqnarray}
where the spin connection is expressed in terms of the Christoffel symbols (there is no torsion in our model) and the tetrads:
\begin{equation}\label{spcon}
\omega_{A\,\tilde{M}\tilde{N}} = h_{\tilde{M}M} \Gamma^M_{AN}h_{\tilde{N}}^N + h_{\tilde{M}M} \partial_A h_{\tilde{N}}^M\,.
\end{equation}

\subsection{Compact extra space}

As is well known, the charge conservation is related to a Lagrangian invariance under a group transformation. In our case, we interpret such invariance as an extra space geometric isometry. Therefore, the extra space metric should be invariant under group transformations. Our analysis is based on metrics that possess U(1) isometry. The simplest example of such sort is the 2-dim apple-like metric
\begin{equation}\label{metr}
d s^{2}=g_{\mu \nu}\left(x^{\alpha}\right) d x^{\mu} d x^{\nu}-r_0^2 e^{2\beta(\theta)} \left(d \theta^{2}+b^{2} \sin ^{2} \theta\,d \varphi^{2}\right).
\end{equation}

Here $g_{\mu \nu} \left(x^{\alpha}\right)$ is 4-dim metric, $r_0$ is the radius of compact 2-dim extra space, coefficient $\beta(\theta)$ defines geometry of extra space, parameterized by azimuthal angle $\theta$, and $b$ is the parameter, related to the deficit or excess of angle $\varphi$ in the extra space.
This metric has been discussed in Ref.~\refcite{Gogberashvili_2007} where the variety of analytical results were obtained. Also, such kind of manifolds have been revealed Refs.~\refcite{2016arXiv160907361R,Gani:2014lka,Rubin:2018ybs,Bronnikov:2020tdo} as the static numerical solutions to the generalized Einstein equations. Some of such solutions can be approximated by metric \eqref{metr} which strongly facilitate the analysis below.

The Dirac equation for a multidimensional fermion field has the form
\begin{equation}\label{dirac}
i h_{\tilde{A}}^{B} \hat\Gamma^{\tilde{A}} \nabla_{B}\Psi = 0\,,\quad  h.c. = 0\,.
\end{equation}
where the tetrad 
\begin{equation}\label{tetrad}
h^B_{\widetilde{A}} =\left(\delta_{\widetilde{\mu}}^{B}\,,\, \frac{1}{r_0 e^{2\beta(\theta)}} \delta_{\widetilde{\theta}}^{B}\,,\, \frac{1}{b r_0 e^{2\beta(\theta)} \sin \theta} \delta_{\widetilde{\varphi}}^{B}\right)\,.
\end{equation}
is associated with the metric \eqref{mettet},\eqref{metr}.

The solution to this equation can be decomposed as:
\begin{equation}\label{decomp}
\Psi\left(X\right) = \sum_{j} Y_j(\theta,\varphi) \Psi_j(x) = \sum_{jl} e^{i l \varphi} \left(\begin{array}{l}{Y^+_{jl}(\theta) \psi_{jl}(x)} \\ {Y^-_{jl}(\theta) \xi_{jl}(x)}\end{array}\right)\,,
\end{equation}
where $\psi_{jl}$ and $\xi_{jl}$ is 4-dim Dirac spinors, and  $Y^+_{jl}$, $Y^-_{jl}$ is complete set of functions (that is the eigenfunctions of the corresponding operator, see \ref{app3} for definition). Number $j$ corresponds to the energy (mass) level of KK-tower and varies from 0 to infinity. We are interested only in massless modes $j=0$.

The integer number $l$ varies in the range $-b < 2l < b$ (see \ref{app3}) and can be associated with the internal angular momentum. This number could describe the triplet of fermions in the ground state if the extra space angle excess varies in the interval $2<b\le4$.

\subsection{Effective low-energy 4-dim action}

Let us obtain the effective 4-dim action starting from its initial form \eqref{action}. Using Kaluza-Klein decomposition \eqref{decomp} (see \ref{app3}, \eqref{dirdecomp},\eqref{massmatr}) and extra space tetrad \eqref{tetrad} we get 4-dim action for the 4-dim fermions (KK-tower):

\begin{eqnarray}
S &=&\int\sqrt{|G|}\,d^{6} X\,\,\overline{\Psi} h_{\tilde{A}}^{B} \hat\Gamma^{\tilde{A}} \nabla_{B} \Psi =\nonumber\\
&=&\int\sqrt{|g|}\sqrt{|k|}\,d^{4}x\,d^2y \left(i \overline{\Psi} h_{\tilde{\mu}}^{\nu} \hat\Gamma^{\tilde{\mu}} \nabla_{\nu}\Psi + i \overline{\Psi} h_{\tilde{m}}^{n} \hat\Gamma^{\tilde{m}} \nabla_{n}\Psi\right)= \nonumber\\
&=&\int\sqrt{|g|}\,d^{4}x\, \sum\limits_{jl}  \left(i \overline{\psi}_{jl} h_{\tilde{\mu}}^{\nu} \gamma^{\tilde{\mu}} \nabla_{\nu}\psi_{jl} + i \overline{\xi}_{jl} h_{\tilde{\mu}}^{\nu} \gamma^{\tilde{\mu}} \nabla_{\nu}\xi_{jl}\, +\right.\nonumber\\
&&\hphantom{\int\sqrt{|g|}\,d^{4}x\, \sum\limits_{jl}\left(\right.} \left.
+\,M_j\overline{\psi}_{jl}\xi_{jl} +  M^\prime_j\overline{\xi}_{jl}\psi_{jl}\right)\,.
\end{eqnarray}

It can be seen that the masses of nonzero modes $j\neq0$ are proportional to the inverse radius of the extra space $M_j,M^\prime_j \sim h^n_{\tilde{m}} \sim 1/r_0$. We are not interested in massive modes $j\neq0$, since they are extremely heavy due to small extra space size $r_0\ll10^{-19}$ cm \cite{Deutschmann2017CurrentDimensions, nikulin2019inflationary}.

Further, we consider only massless modes $\psi_{0l}=\psi_{l},\,\xi_{0l}=\xi_{l}$. As we have indicated above (and in \ref{app3}) there are three independent modes for the ground state $j=0$: $l=-1,0,1$. Thus, the low-energy effective action contains two 4-dimensional fermion triplets:
\begin{eqnarray} \label{psixi}
S&=&\int\sqrt{|g|}\,d^{4} x\, \sum\limits_l \left(i \overline{\psi}_{l} \gamma^{\mu} \partial_{\mu} \psi_{l} + i \overline{\xi}_{l} \gamma^{\mu} \partial_{\mu} \xi_{l}\right)\,,\qquad l=-1,0,+1\,.
\end{eqnarray}

The fermions $\psi_{l}$ and $\xi_{l}$ are indistinguishable from the 4-dimensional point of view, so further we will focus on one of them: $\psi_{l}=\{\psi_{-1},\psi_{0},\psi_{+1}\}$.

\section{Internal momentum as the lepton/baryon number}\label{III}

We will show in this section that fermions $\psi_l$ and $\xi_l$ \eqref{psixi} carry the lepton/baryon number, so it would be reasonable to identify them with the known particles of the Standard Model. Nevertheless, we do not set ourselves such a task, trying to demonstrate the mechanism of lepton/baryon number generation on a toy model.

\subsection{Internal momentum conservation}

The geometric symmetries give rise the conserved currents $\nabla_A J^A=0$ and the conserved charges associated with them \cite{Blagojevic}. Here we consider the compact symmetric extra dimensions. Let the extra{--}spatial metric $k_{mn}$ remains invariant under the infinitesimal isometries $y^m\rightarrow y^m+\xi^m(y)$, where $\xi^m(y)$ is the Killing vector field. According to the Noether’s theorem there is a conserved current associated with such invariance. This current is given by
\begin{equation}
J^A=\frac{\partial L_\text{m}(\chi)}{\partial(\partial_A \chi)}\xi^b\partial_b\chi - \xi^A L_\text{m}(\chi)\,,
\end{equation}
for any matter field $\chi$. Here $L_\text{m}$ is a Lagrangian of matter.
The corresponding conserved number
\begin{equation}\label{numb}
Q =\int J^0 \sqrt{|G|}\, d^3x\,d^2y\,.
\end{equation}

In the lower (vacuum) state, the metric of the extra space \eqref{metr} comes to the unperturbed axially symmetrical configuration: $\beta(y)=\beta(\theta)$ \cite{2017JCAP...10..001B}. The corresponding symmetry is just rotational symmetry along coordinate $\varphi$. It is characterized by the Killing vector field, which in spherical coordinates has the form \cite{nikulin2020formation,gani2015two,banerjee2004gauge}:
\begin{equation}\label{killing}
\xi^m=\left(\begin{array}{l}
     0 \\
     1 
\end{array}\right)\,.
\end{equation}

The conserved number $Q$, corresponding to the Killing vector \eqref{killing}, has the meaning of the total internal angular momentum of the fields along the coordinate $\varphi$:
\begin{equation}\label{QQ}
Q = Q_\Psi + Q_\Phi =\int J^0 \sqrt{|G|}\, d^3x d^2y = \text{const}\,.
\end{equation}
where $J^0$ is time component of  corresponding conserved current for action $S_{\Phi} + S_{\Psi}$. It has the following form:
\begin{eqnarray}\label{J0}
J^0 = J^0_\Psi + J^0_\Phi &=& \frac{\partial\mathcal{L}}{\partial (\partial_0\Psi)} \partial_\varphi\Psi + \frac{\partial\mathcal{L}}{\partial (\partial_0\Phi)} \partial_\varphi\Phi = \nonumber\\
&=& i \overline{\Psi} h_{\tilde{A}}^{0} \Gamma^{\tilde{A}} \partial_{\varphi} \Psi + \partial^0 \Phi\partial_\varphi \Phi\,,
\end{eqnarray}

In the rest of the section \ref{III}, we will consider only the fermion part of \eqref{J0}. In section \ref{IV} we explain why the fermion field alone is not enough to generate the lepton/baryon number, and describe the reason for introducing the scalar field.

\subsection{Effective 4-dim lepton/baryon number}

From the 4-dim observer point of view, the fermion part of conserved current takes the form:
\begin{equation}\label{j0}
j^\mu_\Psi=\frac{\partial\mathcal{L}_4}{\partial (\partial_\mu\psi_l)}t_{ll'}\psi_{l'} = i\sum\limits_l l\,\overline{\psi_{l}}\gamma^\mu\psi_{l}\,,
\end{equation}
\begin{eqnarray}\label{L4T}
\mathcal{L}_4=\sum\limits_l i \overline{\psi}_{l} \gamma^{\mu} \partial_{\mu} \psi_{l}\,,\qquad
t_{ll'}=\int Y_{0l} \, \partial_\varphi Y_{0l'} \sqrt{|k|}\,  d^2 y = il\delta_{ll'}\,,
\end{eqnarray}
where we consider only $\psi_l$ fields in effective Lagrangian $\mathcal{L}_4$ \eqref{psixi}. Right equation in \eqref{L4T} represents the effective 4-dim generator matrix of primordial 6-dim geometric extra space symmetry (for general explanation, see the \ref{app1}).

Associated conserved number $Q_\Psi$ takes the form:
\begin{eqnarray}\label{Q}
Q_\Psi&=&\int j^0_\Psi \,\sqrt{|g|}\,d^3x =\nonumber\\
&=&\int i (\psi_{+1}^\dagger \psi_{+1} - \psi_{-1}^\dagger \psi_{-1})\,\sqrt{|g|}\,d^3x = N_{\psi_{+1}} - N_{\psi_{-1}} =\text{const}\,,
\end{eqnarray}
where the capital letter $ N_\psi$ denotes the 4-dim fermion number of the corresponding particle (the number of particles minus the number of antiparticles), which is positive for an excess of matter and negative for an excess of antimatter.

We see, that internal (extra space) momentum of the multidimensional fermion field is equal to the difference of fermion numbers of its effective 4-dim components with opposite index $l$ and it is conserved. We will interpret the internal momentum $Q_\Psi$ as lepton/baryon number.

\subsection*{Remark on the fermion number}

The fermion action \eqref{action} is invariant under the global phase transformation $\Psi \rightarrow e^{i\theta} \Psi$. This symmetry, in contrast to the one discussed above, is the symmetry of the Lagrangian itself and does not relate to the symmetry of extra space metric. The global invariance is responsible for the conservation of the total multidimensional fermion number
\begin{eqnarray}\label{NN}
N_\Psi&=&\int \Bar{\Psi}\hat\Gamma^0\Psi \sqrt{|G|}\,d^3x\,d^2y = \int \Psi^\dagger\Psi \sqrt{|g|}\sqrt{|k|}\,d^3x\,d^2y=\nonumber\\
&=&\int \sum\limits_l\psi_{l}^\dagger \psi_{l} \,\sqrt{|g|}\,d^3x
= N_{\psi_{+1}} + N_{\psi_{0}} + N_{\psi_{-1}} =
\text{const}\,,
\end{eqnarray}

The particles $\psi_{+1}$ and antiparticles $\psi_{-1}$ have positive lepton (baryon) number $Q_\Psi$, see \eqref{Q}. Therefore we identify these particles with leptons (baryons). A nonzero value $Q_\Psi$ points to a nonzero difference between the leptons (baryons) and antileptons (antibaryons) --- the lepton (baryon) asymmetry.

\section{Lepton/baryon number generation}\label{IV}

In the previous section, we have shown that the effective 4-dimensional fermions $ \psi_l $ carry a conserved number related to the extra space U(1) symmetry which we associate with the lepton/baryon symmetry.
%
At high energies, during the inflation, the geometry of extra space will be disturbed by gravitational waves. The extra space metric has no isometries and, as a consequence, lepton/baryon number \eqref{Q} --- is not conserved \cite{nikulin2020formation}. However, such a process of lepton/baryon number accumulation is forbidden by the quantum nature of fermions (spin-statistics theorem), which cannot be accumulated in the form of coherent field oscillations.

A scalar field can accumulate such a number in the form of coherent oscillations. This is the reason for the introduction of a scalar field in the model. The Lagrangian term for its interaction with fermions provides the transfer of lepton/baryon charge from the scalar particles to fermions. Note that the scalar field also have a lepton/baryon number because it acts in the same extra space. Thus, according to our model the accumulation of the total number \eqref{QQ} takes place only in the scalar field during inflation.

\subsection{Internal momentum of scalar field}

The scalar field charge relates to its internal momentum in the extra space and has the form 
\begin{equation}
Q_\Phi=\int \partial^0\Phi\,\partial_\varphi\Phi \sqrt{|G|}\,d^3x\,d^2y\,
\end{equation}
following from the expression \eqref{J0}.
Total charge
\begin{eqnarray}\label{QQQ}
Q&=&\int J^0 \sqrt{|G|}\,d^3x\,d^2y = \int j^0_\Psi \sqrt{|g|}\, d^3x + Q_\Phi = \nonumber\\
&=&i\int \left(\psi_{+1}^\dagger \psi_{+1} - \psi_{-1}^\dagger \psi_{-1}\right)\sqrt{|g|}\,d^3x + Q_\Phi =\nonumber\\
&=& N_{\psi_{+1}} - N_{\psi_{-1}} + Q_{\Phi}=\text{const}
\end{eqnarray}
includes both the fermion and boson part. This charge is conserved after the inflation is finished and the extra space metric is stabilized. Under these conditions the total fermion number \eqref{NN} is also conserved:
\begin{equation}\label{NNN}
N_\Psi = N_{\psi_{+1}} + N_{\psi_{0}} + N_{\psi_{-1}} =
\text{const}\,.
\end{equation}

Thus, if we consider only massless fermions $ \psi_l $, the field $ \Phi $ can decay into 3 different modes: $l =  -1,0,+ 1 $. Because of the conservation laws \eqref{QQQ}, \eqref{NNN}, the non-zero momentum $ Q_\Phi $ (for example, with a positive sign) can transferred into the $ \psi_{+1} $ mode with the simultaneous generation of a similar number of antifermions $ \psi_0 $. If we identify the lepton (baryon) number with the $ +1 $ mode, we obtain a mechanism for generating lepton (baryon) asymmetry.

\subsection{Decay of scalar field perturbations}

Consider the Yukawa term in action \eqref{scaction} and expand it over the vacuum state of the scalar field $\Phi(y)=\Phi_\text{vac}(y)+\delta\Phi(y)$:
\begin{equation}\label{Yukawa}
S_{\text{int}}=\int \sqrt{|G|}\,d^{6}X \,f\Phi\Bar{\Psi}\Psi = \int \sqrt{|G|}\,d^{6}X \,(f\Phi_\text{vac}\Bar{\Psi}\Psi+f\delta\Phi\Bar{\Psi}\Psi)\,.
\end{equation}

The second term describes the interaction of scalar field perturbations and the fermions. It is the scalar field perturbations $\delta\Phi$ that accumulate the total nonzero internal moment during inflation.


After decomposition and integration over the extra coordinates, the effective 4-dimensional interaction acquires the form 
\begin{equation}\label{sint}
S'_{\text{int}}=\int \sqrt{|G|}\,d^6X\,f\delta\Phi\Bar{\Psi}\Psi=\int \sqrt{|g|}\,d^{4}x\,\left[f A^{lm}\Bar{\psi}_{l}\psi_{m} + f B^{lm}\Bar{\xi}_{l}\xi_{m}\right]\,,
\end{equation}
where functions $A^{lm}(x),B^{lm}(x)$ can be considered as classical  4-dimensional sources proportional to the scalar field fluctuations $\delta\Phi$: \begin{eqnarray}
A^{lm}(x) = \int \delta\Phi(x,\theta,\varphi) Y^+_{l}(\theta)^* Y^+_{m}(\theta)e^{i(l-m)\phi}\,\sqrt{|k|}\,d^2y\,,\nonumber\\
B^{lm}(x) = \int \delta\Phi(x,\theta,\varphi) Y^-_{l}(\theta)^* Y^-_{m}(\theta)e^{i(l-m)\phi}\,\sqrt{|k|}\,d^2y\,.
\end{eqnarray}

The vertices corresponding to the interaction \eqref{sint} describe the decay of the scalar field condensate $\delta\Phi$ into effective 4-dimensional fermions $\psi_l, \xi_l$:

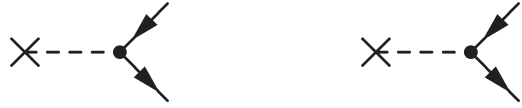
\vspace{0.4cm}
\begin{fmffile}{d1}
\begin{minipage}{0.55\textwidth}
\begin{center}
\begin{fmfgraph*}(60,37)
    \fmfleft{i1}
    \fmfright{o1,o2}
    \fmfv{label=$A^{lm}$,l.a=180,l.d=.1w}{i1}
    \fmfv{label=$\psi_{l}$,l.a=0,l.d=.05w}{o1}
    \fmfv{label=$\overline{\psi}_m$,l.a=0,l.d=.05w}{o2}
    \fmfv{label=$f$,l.a=110,l.d=.1w}{w1}
    \fmf{fermion}{o2,w1,o1}
    \fmf{dashes}{i1,w1}
    \fmfdot{w1}
    \fmfv{decor.shape=cross}{i1}
\end{fmfgraph*}
\end{center}
\end{minipage}
\begin{minipage}{0.2\textwidth}
\begin{fmfgraph*}(60,37)
    \fmfleft{i1}
    \fmfright{o1,o2}
    \fmfv{label=$B^{lm}$,l.a=180,l.d=.1w}{i1}
    \fmfv{label=$\xi_l$,l.a=0,l.d=.05w}{o1}
    \fmfv{label=$\overline{\xi}_{m}$,l.a=0,l.d=.05w}{o2}
    \fmfv{label=$f$,l.a=110,l.d=.1w}{w1}
    \fmf{fermion}{o2,w1,o1}
    \fmf{dashes}{i1,w1}
    \fmfdot{w1}
    \fmfv{decor.shape=cross}{i1}
\end{fmfgraph*}
\end{minipage}
\end{fmffile}

\subsection{Higgs mechanism}\label{Higgs}

We assumed zero masses of the fermions from the beginning. In theories with extra dimensions, there are a variety of ways to give mass to the fermions. For example, in Ref.~\refcite{neronov2002fermion} the emergence of masses and mixing is associated with the presence of angular momentum in the extra space.

For our purposes, there is already a scalar field that naturally induces the Higgs mechanism.
Consider the first term in \eqref{Yukawa}:
\begin{eqnarray}\label{effmass}
S_\text{int} &=& \int \sqrt{|G|}\,d^{6}X\, f\Phi_\text{vac}\overline{\Psi}\Psi
=\nonumber\\
&=&\int \sqrt{|g|} d^{4}x \Bigg[\left(\int\sqrt{|k|}\,d^{2}y\, \Phi_\text{vac}(y)(Y^+_l)^*Y^+_m\right)\overline{\psi}_l\psi_m +\nonumber\\
&&\hphantom{\int \sqrt{|g|}\, d^{4}x}+\left(\int\sqrt{|k|}\,d^{2}y\, \Phi_\text{vac}(y)(Y^-_l)^*Y^-_m\right)\overline{\xi}_l\xi_m\Bigg] =\nonumber\\
&=&\int \sqrt{|g|}\,d^{4}x\, (M^+_{lm}\overline{\psi}_l\psi_m + M^-_{lm}\overline{\xi}_l\xi_m)\,.
\end{eqnarray}
In \eqref{effmass}, we used the expansion of fermions in terms of eigenfunctions \eqref{decomp}. The vacuum state $\Phi_\text{vac}$ is a homogeneous in 4 dimensions and stationary solution of the Klein-Gordon equation:
\begin{equation}
\Box_M\Phi_\text{vac}(X)+V'(\Phi_\text{vac})=\Box_K\Phi_\text{vac}(y)+V'(\Phi_\text{vac})=0
\end{equation}
where $\Box_M, \Box_K$ is the multidimensional and extradimensional d'Alembert operators, respectively. Due to the rotational symmetry of the vacuum state, $\Phi_\text{vac}(y)=\Phi_\text{vac}(\theta)$, which leads to the diagonalization of the matrices $M^{\pm}_{lm}\sim\int e^{i(l-m)\varphi}d\varphi \sim \delta_{lm}$.

The structure of eigenfunctions according to the \ref{app1} is as follows
\begin{align}
Y^+_{+1}=Y^-_{-1}\sim \cot(\theta/2)\quad&\implies\quad m_{\phi_{+1}}=m_{\xi_{-1}}=m_+\,,\nonumber\\
Y^-_{+1}=Y^+_{-1}\sim \tan(\theta/2)\quad&\implies\quad m_{\xi_{+1}}=m_{\psi_{-1}}=m_-\,,\nonumber\\
Y^-_{0}=Y^+_{0}\sim 1\qquad\qquad&\implies\quad m_{\phi_{0}}=m_{\xi_{0}}=m_0\,.
\end{align}

Thus, 3 different masses arise, determined by the dependence of the eigenfunction on the azimuthal angle $\theta$. In addition, due to the azimuthal asymmetry of the apple-like extra space, a hierarchy arises $m_+<m_0<m_-$. 


\section{Conclusion}

In this paper, we study a possible mechanism for generating the lepton/baryon number and its subsequent conservation based on the extra space evolution. The internal momentum is associated with the number.

Our first articles \cite{Rubin:2018ybs} were devoted to the accumulation of charge in extradimensional scalar field excitations. It is assumed that the metrics of manifolds initially do not have symmetries. The symmetries appear later due to the entropy growth \cite{Kirillov:2012gy,nikulin2020formation}.
Here we continue the research and study the way of the charge transition from the scalar field to the leptons/baryons. 

We consider a toy multidimensional field model that allows us to establish how the internal angular momentum is accumulated in the scalar field and finally transferred to the fermions. 

The developed model demonstrates the consequences of the symmetry restoration in Kaluza-Klein theories. The mechanism itself is quite general and can be applied to any other internal symmetry (for example, baryonic, electroweak, etc.). Nevertheless, a realistic implementation of this mechanism, which makes it possible to obtain the known SM leptons/baryons in the low-energy limit, is questionable. The research in this direction will be the subject of our further studies.

\section*{Acknowledgments}

This research was funded by the Ministry of Science and Higher Education of the Russian Federation, Project ``Fundamental properties of elementary particles and cosmology'' N 0723-2020-0041.
The work of S.R. has been supported by the Kazan Federal University Strategic Academic Leadership Program.

\appendix

\section{KK-decomposition of 6-dim fermion}\label{app3}

Let us start with the 6-dimensional Dirac equation
\begin{equation}
i h_{\tilde{A}}^{B} \hat\Gamma^{\tilde{A}} \nabla_{B}\Psi = 0\,,\quad  h.c. = 0\,.
\end{equation}
Here $\Psi$ is the 8-components spinor and flat gamma matrices are chosen in the form
\begin{equation}
\hat\Gamma_{\nu}=\left(\begin{array}{cc}{\gamma_{\nu}} & {0} \\ {0} & {-\gamma_{\nu}}\end{array}\right), \quad \hat\Gamma_{\theta}=\left(\begin{array}{cc}{0} & {-1} \\ {1} & {0}\end{array}\right), \quad \hat\Gamma_{\varphi}=\left(\begin{array}{cc}{0} & {i} \\ {i} & {0}\end{array}\right)\,.
\end{equation}

For the space that is a direct product $M_4 \times K$, we can choose coordinates in such a way that the indices of the 4-dimensional and extra-dimensional parts of the metric are separated:
\begin{equation}
G_{m\nu}=0\quad\implies\quad h^{m}_{\tilde{\nu}}=h^{\mu}_{\tilde{n}}=0\,.
\end{equation}
It strongly simplifies the form of the Christoffel symbols and spin connection
\begin{eqnarray}
\Gamma^a_{\mu\nu}=\Gamma^\alpha_{m\nu}&=&\Gamma^a_{m\nu}=\Gamma^\alpha_{mn}=0\,,\nonumber\\
\omega_{a \tilde\mu \tilde\nu}=\omega_{\alpha \tilde{m} \tilde\nu}&=&\omega_{\alpha \tilde{m} \tilde\nu}=\omega_{\alpha \tilde{m} \tilde{n}}=0\,.
\end{eqnarray}

Thus, the Dirac equation can be completely split into 2 terms:
\begin{equation}\label{effdir}
i h_{\tilde{A}}^{B} \hat\Gamma^{\tilde{A}} \nabla_{B}\Psi = i h_{\tilde{\mu}}^{\nu} \hat\Gamma^{\tilde{\mu}} \nabla_{\nu}\Psi + i h_{\tilde{m}}^{n} \hat\Gamma^{\tilde{m}} \nabla_{n}\Psi = 0\,.
\end{equation}
The second term in \eqref{effdir} acts as the mass operator of effective 4-dimensional fields. We now expand the 6-dimensional fermion field in terms of the complete orthogonal system of functions of extra coordinates $\Psi(x,y)=\sum\limits_j \Psi_j(x) Y_j(\theta,\phi)$:

\begin{equation}\label{dirdecomp}
\sum\limits_{j=1}^\infty i h_{\tilde{\mu}}^{\nu} \hat\Gamma^{\tilde{\mu}} \nabla_{\nu}\Psi_j\,Y_j + \sum\limits_{j=1}^\infty \left[i h_{\tilde{m}}^{n} \hat\Gamma^{\tilde{m}} \partial_{n} + \frac{i}{4} h_{\tilde{m}}^{n} \omega_{n \tilde{a} \tilde{b}} \hat\Gamma^{\tilde{m}} \hat\Gamma^{\tilde{a}} \hat\Gamma^{\tilde{b}}\right]Y_j\,\Psi_j = 0\,.
\end{equation}
A complete system of functions $\{Y_j\}$ is the eigenfunctions of the operator enclosed in brackets in \eqref{dirdecomp}. These eigenfunctions correspond to the states with a definite mass \cite{witten1981search}.
It can be shown that the product  $\hat\Gamma^{\tilde{A}}\hat\Gamma^{\tilde{M}}\hat\Gamma^{\tilde{N}}$ has a diagonal structure for 4-dim indices and antidiagonal for extra space indices:

\begin{table}[ph]
\tbl{Block structure of gamma matrices and their products for 4 + 2 dimensions.}
{\begin{tabular}{@{}cc|cc@{}} \toprule
$\tilde{M}$ &
$\hat\Gamma^{\tilde{M}}$ & $\tilde{A}\tilde{M}\tilde{N}$ & 
$\hat\Gamma^{\tilde{A}}\hat\Gamma^{\tilde{M}}\hat\Gamma^{\tilde{N}}$ \\ \colrule
$\tilde{\mu}$ &
$\begin{bmatrix}
4\times4 & 0\\
0 & 4\times4
\end{bmatrix}$ &
$\tilde{\alpha}\tilde{\mu}\tilde{\nu}$ & $\begin{bmatrix}
4\times4 & 0\\
0 & 4\times4
\end{bmatrix}$ \\ & & & \\
$\tilde{m}$ & $\begin{bmatrix}
0 & 4\times4\\
4\times4 & 0
\end{bmatrix}$ &
$\tilde{a}\tilde{m}\tilde{n}$ & $\begin{bmatrix}
0 & 4\times4\\
4\times4 & 0
\end{bmatrix}$\\
\botrule
\end{tabular}\label{gblock}}
\end{table}

Thus, the upper and lower 4-components of the spinor  $\Psi_j(x)=\{\psi_j(x),\xi_j(x)\}$ are separated. The second term in \eqref{dirdecomp} plays the role of the mass term mixing the 4-spinors $\psi_j(x),\xi_j(x)$:
\begin{equation}
\begin{bmatrix}
i h_{\tilde{\mu}}^{\nu} \gamma^{\tilde{\mu}} \nabla_{\nu} & 0\\
0 & - i h_{\tilde{\mu}}^{\nu} \gamma^{\tilde{\mu}} \nabla_{\nu}
\end{bmatrix}
\begin{bmatrix}
\psi_j\\
\xi_j
\end{bmatrix} + 
\begin{bmatrix}
0 & M_j\\
M^\prime_j & 0
\end{bmatrix}
\begin{bmatrix}
\psi_j\\
\xi_j
\end{bmatrix} = 0\,,\quad\text{where}\nonumber
\end{equation}
\begin{equation}\label{massmatr}
\left[i h_{\tilde{m}}^{n} \hat\Gamma^{\tilde{m}} \partial_{n} + \frac{i}{4} h_{\tilde{m}}^{n} \omega_{n \tilde{a} \tilde{b}} \hat\Gamma^{\tilde{m}} \hat\Gamma^{\tilde{a}} \hat\Gamma^{\tilde{b}}\right]Y_j=
\begin{bmatrix}
0 & M_j\\
M^\prime_j & 0
\end{bmatrix}Y_j\,.
\end{equation}

We are interested in finding massless modes with non-zero angular momentum. There is strict theorem \cite{ABRIKOSOV2002, Camporesi1996} that there are no zero fermion modes on the $S^2$ sphere. Nevertheless, there are ways to violate this theorem using exotic geometry, for example, using the warp factor with special boundary conditions \cite{Kokado_2015}. Here we will briefly summarize the approach based on the extra metric with the angle excess \cite{Gogberashvili_2007}. To find the massless $Y_0$ modes, we have to solve the equation \eqref{massmatr} with zero right-hand side: 
\begin{equation}
\left[i h_{\tilde{m}}^{n} \hat\Gamma^{\tilde{m}} \partial_{n} + \frac{i}{4} h_{\tilde{m}}^{n} \omega_{n \tilde{a} \tilde{b}} \hat\Gamma^{\tilde{m}} \hat\Gamma^{\tilde{a}} \hat\Gamma^{\tilde{b}}\right]Y_0=0\,.
\end{equation}

To obtain an analytical solution to the equation above, we expand $Y_0(\theta,\phi)$ in terms of the complete orthogonal system of functions along the polar coordinate $Y_{0}(\theta,\phi)=\sum\limits_l e^{il\phi} Y_{0l}(\theta)$. As a result, for the apple-like metric \eqref{metr} we obtain two equations for $Y^{\pm}_{0l}(\theta)$ (from the top ($-$) and bottom ($+$) blocks of gamma matrices, respectively):

\begin{eqnarray}
Y^{-}_{0l} \left( (b \sin \theta+1)(\beta_{\theta } \sin \theta + \cos \theta)-2l/b\right)+2 \sin \theta\,Y^{-\prime}_{0l}&=&0\,,\nonumber\\
Y^{+}_{0l} \left( (b \sin \theta+1)(\beta _{\theta } \sin \theta + \cos \theta)+2l/b\right)+2 \sin \theta\,Y^{+\prime}_{0l}&=&0\,.
\end{eqnarray}

The analytical solution of these equations gives:
\begin{equation}
Y^\pm_{0l} = C e^{-\frac{1}{2}\int_0^\theta \beta_{\theta'}(b\sin\theta'+1)d\theta'} \,\frac{\tan^{\pm l/b}\,\theta/2}{\sqrt{\sin\theta}}\,.
\end{equation}

Consider an undeformed metric $\beta = 0$ with the excess $b$ of the angle. It turns out that the scalar square of the eigenfunction for massless modes $Y_0$ is finite only for certain $l$:
\begin{align}\label{normcond}
\langle Y_0|Y_0 \rangle=&\int|Y_0|^2\,b\sin\theta\,d\theta \,d\phi=\nonumber\\
=& \int\left(|Y_{0l}^+|^2+|Y_{0l}^-|^2\right) \, 2\pi b \sin\theta\,d\theta<\infty\,,\quad\text{for}\quad
-1<\frac{2l}{b}<1\,.
\end{align}
There is only one massless mode $l = 0$  for $b = 1$ (which is intuitively clear). If the space has angle excess, for example $b = 3$, splitting occurs and three modes $ l = -1,0,1 $ are finite.

If the metric of the extra space is deformed $\beta(x) \neq 0 $, then the coefficient $ C [\beta] $ appears in the condition \eqref{normcond}. This coefficient is a functional of the extra space geometry:
\begin{equation}\label{coefflimit}
-1< C[\beta]\frac{2l}{b}<1\,.
\end{equation}
Therefore we can achieve the splitting effect even without introducing the global geometry parameter $ b \neq 1 $, using only the coefficient $ C [\beta] $.

\section{Currents and symmetries in D dimensions}\label{app1}

Let us show that the extra metric symmetry looks as an internal symmetry of the 4-dimensional fields for the 4-dimensional observer. This is manifested in the charge conservation.

Geometric symmetry is the invariance of the metric under the infinitesimal coordinate transformations: $X'_A=X_A+\xi_A^i(X) \,\varepsilon_i$, where $\xi_A^i(X)$ is Killing vector fields and $\varepsilon_i\rightarrow0$. The Noether's theorem leads to a general formula for currents conserved due to geometric symmetry (hereinafter called "geometric currents"):
\begin{equation}\label{geom_curr}
J^A_i=\frac{\partial \mathcal{L}}{\partial(\partial_A\phi)}\hat{T}_i\phi - \xi^A_i \mathcal{L}\,,\quad\text{where}\quad \hat{T}_i=\xi^A_i\partial_A\,.
\end{equation}

For example, substituting the Killing translation vectors for the Minkowski space, we obtain the energy-momentum tensor, and substituting the rotational ones, we obtain the angular momentum tensor.

In the case of internal (isotopic) symmetry, Lagrangian is invariant under the infinitesimal field transformations: $\phi'_n=\phi_n+(t_i)^n_m\phi^m \,\varepsilon^i$, where $(t_i)^n_m $ are the generators of the internal symmetry group, $\phi^n$ is the field multiplet that transforms according to a representation of this group and $\varepsilon^i\rightarrow0$. The general formula for the currents conserved due to this symmetry (hereinafter referred to as "charge currents"): \begin{equation}\label{internal_curr}
j^A_i=\frac{\partial \mathcal{L}}{\partial(\partial_A\phi^n)}(t_i)^n_m\phi^m\, ,
\end{equation}

For example, substituting generators for SU(3) group and the quark triplets, we obtain the color charged currents.

The aim of this appendix is to relate \eqref{internal_curr} and \eqref{geom_curr}. It follows from Noether's theorem that the current \eqref{geom_curr} satisfies $\nabla_A J^A_i=0$. Let's integrate it over the extra space and split its 4-dim and extra-dim parts:
\begin{eqnarray}
0&=&\int \nabla_A J^A_i\, \sqrt{|k|}\,d^dy = \int \left(\nabla_\alpha J^\alpha_i+\nabla_a J^a_i \right)\, \sqrt{|k|}\,d^dy =\nonumber\\
&=&\nabla_\alpha \left(\int J^\alpha_i \sqrt{|k|}\,d^dy \right) + \int \partial_a \left(J^a_i\sqrt{|k|}\right)\,d^dy = \nabla_\alpha j_i^\alpha\,,
\end{eqnarray}
where we rename $\int J^\alpha_i\sqrt{|k|}\, d^dy = j_i^\alpha$. Thus, $j_i^\alpha$ represents the 4-dimensional conserved current because $\nabla_\alpha j_i^\alpha = 0$. Here we use the assumption that the multidimensional space $M=M_4\times K$ is the direct product of the 4-dimensional space $M_4$ with metric $g_{\alpha\beta}$ and the compact extra space $K$ with metric $k_{ab}$. Therefore $\sqrt{|G(x,y)|}=\sqrt{|g(x)|}\sqrt{|k(y)|}$.

Consider only that part of symmetries (and, accordingly, Killing vectors), which corresponds to the extra dimensions $ \xi^A_i = (0, \xi^a_i) $.
According to \eqref{geom_curr}, we write:
\begin{eqnarray}\label{reduce}
j_i^\alpha& = &\int J^\alpha_i\sqrt{|k|}\,d^dy = \int\frac{\partial \mathcal{L}}{\partial(\partial_\alpha\phi)}\hat{T}_i \phi\sqrt{|k|}\,d^dy =\nonumber\\
& = &\int\frac{\partial \mathcal{L}}{\partial(\partial_\alpha\phi^nY_n)}\hat{T}_i Y_m \phi^m\sqrt{|k|}\,d^dy = \int\frac{\partial \mathcal{L}_4}{\partial(\partial_\alpha\phi^n)}Y_n\hat{T}_i Y_m\phi^m\sqrt{|k|}\,d^d y =\nonumber\\
& = &\frac{\partial \mathcal{L}_4}{\partial(\partial_\alpha\phi^n)}\left(\int Y_n\hat{T}_i Y_m\sqrt{|k|}\,d^dy\right) \phi^m
= \frac{\partial \mathcal{L}_4}{\partial(\partial_\alpha\phi^n)} (t_i)^n_m \phi^m\,,
\end{eqnarray}
where $(t_i)^n_m=\int Y_n\hat{T}_i Y_m\sqrt{|k|}\,d^dy$, and the relation for the effective 4-dim Lagrangian $\mathcal{L}_4$ is used:

\begin{eqnarray}\label{poln2}
\frac{\partial \mathcal{L}(x,y)}{\partial(\partial_\alpha\phi^nY_n)} & = & \int\frac{\partial \mathcal{L}(x,\tilde{y})}{\partial(\partial_\alpha\phi^m \tilde{Y}_m)}\delta^d(y-\tilde{y})\,\sqrt{|k|}\, d^d\tilde{y} = \int\frac{\partial \mathcal{L}(x,\tilde{y})}{\partial(\partial_\alpha\phi^m \tilde{Y}_m)}\tilde{Y}_n Y_n\,\sqrt{|k|}\, d^d\tilde{y} =\nonumber\\
& = &\int\frac{\partial \mathcal{L}(x,\tilde{y})}{\partial(\partial_\alpha\phi^m \tilde{Y}_m)}
\frac{\partial(\partial_\alpha\phi^m \tilde{Y}_m)}{\partial(\partial_\alpha\phi^n)} Y_n\,\sqrt{|k|}\, d^d\tilde{y} = \int\frac{\partial \mathcal{L}(x,\tilde{y})}{\partial(\partial_\alpha\phi^n)} Y_n\,\sqrt{|k|}\,d^d\tilde{y} =\nonumber\\
& = &  \frac{\partial \left( \int \mathcal{L}(x,\tilde{y})\,\sqrt{|k|}\, d^d\tilde{y}\right)}{\partial(\partial_\alpha\phi^n)} Y_n = \frac{\partial \mathcal{L}_4}{\partial(\partial_\alpha\phi^n)} Y_n\,,
\end{eqnarray}
where the completeness property of the system of eigenfunctions is used: $Y_n(\tilde{y}) Y_n(y) = \delta^d(y - \tilde{y})$.

Thus, for a 4-dimensional observer, the geometric currents of the multidimensional fields, acting in the extra space, look like the internal (isotopic) charge currents of the effective 4-dimensional fields:
\begin{eqnarray}
\ \ \ \nabla_A J_i^A = 0,&\quad& J^A_i=\frac{\partial \mathcal{L}}{\partial(\partial_A\phi)}\hat{T}_i\phi - \xi^A_i \mathcal{L}\,,\nonumber\\
&&\hat{T}_i=\xi^A_i\partial_A,\quad \xi^A_i=(0,\xi^a_i)
\end{eqnarray}
\begin{eqnarray}
\implies \nabla_\alpha j_i^\alpha = 0,&\quad&  j^\alpha_i=\frac{\partial \mathcal{L}_4}{\partial(\partial_\alpha\phi^n)}(t_i)^n_m\phi^m\,,\nonumber\\
&&(t_i)^n_m=\int Y_n\hat{T}_i Y_m\sqrt{|k|}\,d^dy\,.
\end{eqnarray}

Therefore the corresponding effective 4-dimensional numbers in the comoving volume are conserved:
\begin{equation}
Q_i=\int j^0_i\sqrt{|g|}\,d^3 x = \text{const}\,.
\end{equation}

\bibliographystyle{ws-ijmpd}
\bibliography{biblio}

\end{document}